# Mars before the space age

BARRIE W. JONES

**Astronomy Group, The Open University, Milton Keynes, MK7 6AA, UK**

Email: b.w.jones@open.ac.uk     Phone: +44 1908 653229     Fax: +44 1908 654192

**(SHORT TITLE: Mars pre space age)**

**Abstract**

Mars has surely been scrutinised since the dawn of humankind. Its appearance every couple of years like a drop of blood in the sky, led to warlike attributes in the ancient world. In the 16$^{th}$ century Tycho Brahe made accurate observations of the position of Mars, that enabled Johannes Kepler to obtain his first two laws of planetary motion. These in turn were explained by Newton's laws of motion and gravity. In the 17$^{th}$ century the first telescope observations were made, but Mars is small, and very little surface detail could be discerned.

Throughout the 18$^{th}$ and 19$^{th}$ centuries telescopes improved, revealing many dark areas on the red tinted surface. During the close opposition of 1877 sufficient detail could be seen that enabled Giovanni Schiaparelli to announce that he could see about 40 *canali* on Mars. This led to the saga of the canals of Mars, finally laid to rest in 1971 when Mariner 9 made observations from martian orbit showing that the *canali*/canals do not exist.

Belief that there was life on Mars was widespread in the nineteenth century. However, the majority of astronomers never believed in martian *intelligence*. Least controversial was the view that the dark areas were some form of plant life. This view persisted until Mariner 4 flew past Mars in 1965 and discovered a far thinner atmosphere than previously thought. This was a low point, with impact craters dominating the images. It was Mariner 9 that revealed much more promising landscapes, including volcanic features, and others indicating that water had flowed across the surface, particularly when Mars was young. Thus, the contemporary era of Mars exploration began.

Our picture of Mars today is not only much more complete than that before Mariner 4, in several ways it is quite different. But the belief that there might be life on Mars persists – subsurface life cannot be ruled out, and failing that, there might be ancient fossils on Mars.

**Mars – our current knowledge**

Before I outline what was known about Mars before the flyby of Mariner 4 in 1965, here is a very brief summary of our current knowledge of Mars, with an emphasis on features relevant to the possibility of martian life.

Mars is a small world, 53% the diameter of the Earth. Like the Earth it consists mainly of an iron rich core overlain by a silicate rich mantle, topped by a thin crust rich in silicates and in more volatile materials including water. Its orbit has a semimajor axis of 1.524 AU and an eccentricity of 0.0936. For the Earth these values are 1.000 AU (by definition) and 0.0167 respectively. Mars is our neighbour, being the next planet out from the Sun.

Mars has a thin atmosphere consisting largely of $CO_2$, with a column mass of only $0.015 \times 10^4$ kg m$^{-2}$ (for the Earth the value is $1.07 \times 10^4$ kg m$^{-2}$). This thin atmosphere, the sparse cloud cover, and the distance of Mars from the Sun, have made it a cold world. On a really good day in the tropics the temperature can reach up to around 20 °C, but it plunges to –100 °C or even lower at night. The surface has a reddish tint strewn with darker areas (Figure 1). It divides into two roughly equal areas, divided by a line inclined at nearly 40° with respect to the Equator. The northerly hemisphere shows considerable evidence of geological activity, perhaps confined to the past, and a corresponding low density of impact craters. The southerly hemisphere, being older, is much more impact scarred, and displays little evidence of geological activity. It does, however, bear features that seem to have been carved by liquid water (e.g. the channels in Figure 13), that suggest Mars was probably a much warmer, wetter place early in the first billion years or so of its 4.6 billion year history. Today, water has only been seen in solid form, in the residual north polar cap, in frosts that form at night, and as thin clouds of ice crystals, though it is probably present as liquid in places beneath the surface. Evidence for such subsurface liquid water is provided by narrow gulleys on slopes, some of which seem to have been carved in the last few years (Figure 2).

> Figure 1    Hubble Space Telescope images of the whole surface of Mars, in August 2003 when the planet was particularly close to the Earth. The white region at the bottom of each image is the residual south polar cap ($CO_2$ frost and snow). The upward pointing dark triangular feature in the left image is Syrtis Major. (NASA/STScI, J Bell & M Wolff)

> Figure 2    Gullies in the south facing wall of the channel Nirgal Vallis, imaged by Mars Global Surveyor in 2000. Frame width 2.3 km. (NASA/JPL/Malin Space Science Systems)

$CO_2$, also condenses to form clouds and haze. It also condenses on to the surface, and accounts for the spread of each polar cap in that hemisphere's winter – Mars's axial inclination is similar to that of the Earth (25.2° and 23.4° respectively), and so both planets suffer comparable seasonal changes in insolation, though as the martian "year" is 1.88 of our years, each season lasts considerably longer than on Earth.

This is a very brief outline of our knowledge of Mars today. What about the past, and in particular before the advent in 1965 of the exploration of Mars by spacecraft?

**Mars before telescope observations**

Every 778 days on average, Mars, as seen from the Earth, moves to lie in the opposite direction from the Sun. Around such oppositions it becomes one of the brightest objects in the night sky. The regular motion among the stars of such a bright object must have drawn the attention of our ancestors well before they could record what they saw. From the dawn of written history a few millennia BC, Mars, and the four other visible planets that move among the stars (Mercury, Venus, Jupiter, and Saturn), were considered to be gods by many cultures. For example, the Greek name for Mars is Ares, the god of savage war, or bloodlust, fitting for the red tint of Mars, like a drop of blood in the sky. The Romans also associated the planet with war – Mars is the Roman god of war.

Over the centuries many observations were made of the regular variation of the position of Mars with respect to the background of stars. Of great consequence is the particularly accurate observations made between 1576 and 1597 by the Danish nobleman and astronomer Tycho Brahe (1546-1601, Figure 3). His young assistant the German astronomer Johannes Kepler (1571-1630, Figure 3), who joined him in 1600, spent a long time trying to understand the path of Mars, and in 1609 he announced his first two laws of planetary motion.

*First law*    Each planet moves around the Sun in an ellipse, with the Sun at one focus, the other being empty.

*Second law*   The line from the planet to the Sun sweeps out equal areas in equal times.

Figure 4(a) shows an orbit with a much greater eccentricity than that of Mars, to make Kepler's laws pictorially clear. Figure 4(b) shows the orbit of Mars, with that of the Earth for comparison. It was Mars that led Kepler to his first two laws, because though Mercury's orbit has an even greater eccentricity than that of Mars, it orbits much closer to the Sun than the Earth and so is impossible to observe over much of its orbit. Venus, Jupiter, and Saturn have much less eccentric orbits, and so departures from circular form are harder to discern. Additionally, Jupiter and Saturn have much longer orbital periods than Mars, so have to be observed for much longer to cover a complete orbit. (Kepler's third law also made use of Mars's orbit, and states that the square of the sidereal period of a planetary orbit divided by the cube of the semimajor axis has the same value for all the planets in the Solar System.)

Figure 3 Monument in Prague to Tycho Brahe (left) and Johannes Kepler (right).

Figure 4 (a) An eccentric ellipse, which illustrates Kepler's first two laws of planetary motion. The semimajor axis of the orbit is $a$ and its eccentricity $e$. The shaded areas are equal and are swept out in equal times. (b) The orbits of Mars and the Earth. The point p on each orbit is the position of perihelion (the point closest to the Sun). The point a on the Earth's orbit is the position of aphelion (the point furthest from the Sun).

Kepler's laws were a change in paradigm. Before, and for some time afterwards, it was believed, in both of the geocentric and heliocentric systems, that planets moved in circular paths, which had by Kepler's time become an unwieldly set of circles, several per planet, involving epicycles (e.g. Pecker 2001). Furthermore, Kepler's laws can be explained by the laws of motion and law of gravity of the British philosopher Isaac Newton (1643-1727, Murray & Dermott 1999). This was an early success for Newton's laws.

**Early telescope observations**

The exploration of Mars itself, as opposed to its orbit, began in 1609 when the Italian natural philosopher Galileo Galilei (1564-1642, Figure 5(a)) turned the newly invented telescope to the skies. He saw Mars as a disc, rather than as a point of light in the sky. At that time this was an important discovery in itself, because it suggested that Mars was a world. But it was not until the opposition of 1659 that the first clear drawings of surface markings were made. These are by the Dutch astronomer Christiaan Huygens (1629-1695, Figure 5(b)) who drew a large roughly triangular surface feature, dark on the otherwise red tinted surface (Figure 6). There is little doubt that this is the dark feature subsequently named Syrtis Major (compare Figure 1). Moreover, Huygens was able to follow it night by night and recorded in his diary on 01 December 1659 that (Huygens, in F Terby 1873)

"The rotation of Mars seems to take 24 terrestrial hours like that of the Earth."

He was very nearly right – the rotation period of Mars with respect to the Sun, the "day" on Mars, is 24h 39.6m, just 39.6m longer than our day.

>   Figure 5         (a) Galileo Galilei in 1636, by Justus Sustermans. (b) Christiaan Huygens as a young man.

>   Figure 6         Syrtis Major, sketched by Christiaan Huygens in 1659. North is at the top.

There were many telescope observations of Mars after Huygens. Of particular note are the drawings by the German amateur astronomer Johann Hieronymus Schröter (1745-1816, Schröter 1881). He, and later observers, recorded changes in the shape, extent, and in the contrast of the dark regions against the comparatively light red tint of the rest of the martian surface. Some of these changes followed the seasons. By 1877 many dark markings had been mapped on the surface and changes in these markings had been extensively studied. Most astronomers believed that the dark areas were seas and the light areas continents. A few astronomers believed that the dark areas were plant life, perhaps filling basins that had once been seas.

It had long been known that Mars possessed white polar caps and that these advanced and retreated with the seasons, being largest at the end of winter and smallest at the end of summer. It was known, from observations of surface features as Mars spins rotates, that the inclination of the rotation axis was about 24° and therefore that the seasonal variations in solar radiation are comparable with those on Earth, though the greater eccentricity of the martian orbit significantly enhances the seasonal variations in the southern hemisphere where midsummer occurs soon after perihelion. This is why the southern polar cap advances and retreats with the seasons to a far greater extent than the northern polar cap. The caps were widely thought to consist of water, condensed as ice and snow as on the Earth.

There was ample evidence by 1877 for a martian atmosphere. The polar caps could not exist without an atmosphere – they would rapidly sublime and the water would escape to space, and this would also apply to any open bodies of liquid water. Moreover, as early as 1809 yellow clouds had been observed, sometimes widespread, and by 1858 small white clouds had been seen.

There were no measurements of temperature at the martian surface, but calculations, laced with a heavy dose of speculation, gave values not much lower than the Earth's surface temperatures.

On such an apparently warm moist planet the possibility of life seemed well worth entertaining, and indeed was widely believed to be present (e.g. Camille Flammarion 1842-1925, 1862). By 1870 several schemes had been proposed for signalling to the inhabitants of Mars from the Earth (Drake and Sobel 1993).

This was the position as the opposition of 05 September 1877 gradually approached. Advances in optics and in scientific equipment in general meant that the astronomical community was considerably better prepared than it had been for earlier oppositions. This was a favourable opposition, with Mars not far from perihelion (Figure 4(b)). It is not hard to imagine the excitement mounting as Mars brightened and its angular diameter grew.

*Oppositions of Mars*

Figure 4(b) shows the orbits of Mars and the Earth (these orbits lie almost in the same plane). You can see that the distance between the two orbits varies considerably, largely due to the comparatively large eccentricity of Mars's orbit. Every 780 days on average the Earth overtakes Mars "on the inside lane" and when, as seen from the Earth, Mars and the Sun are in opposite directions, Mars is said to be in opposition. Around opposition Mars and the Earth have their closest approach since the previous "overtaking". The opposition distance of Mars from the Earth varies considerably, depending on where Mars and the Earth are in their orbits at opposition. If Mars is then near its closest to the Sun (perihelion) the distance is comparatively small, and the opposition is called favourable. Unfavourable oppositions are with Mars near aphelion. The opposition distance varies from 55.7-101 million km, and the corresponding angular diameter of Mars varies from 25.1-13.8 arcsec. Favourable oppositions occur roughly every 15 years. Figure 1 shows Mars at a very favourable opposition – our neighbour can hardly get any closer.

**The 1877 opposition and its legacy**

Throughout the weeks surrounding the 1877 opposition the Italian astronomer Giovanni Virginio Schiaparelli (1835-1910) scrutinised Mars visually at the Brera Observatory in Milan where he was director. He used a 220 mm aperture refractor, a large telescope at that time.

His maps of Mars were the best yet and we still use the names he gave to the various dark features, including Syrtis Major. But he is best remembered for about 40 fine lines that he drew crossing the bright red areas, *canali* as he called them. The Italian word *canali* means grooves but drop the "i" at the end and you have a sensation! Description becomes interpretation and in a climate that considered life on Mars at least a reasonable possibility it was not entirely ridiculous to imagine that intelligent life was present and that it had built canals.

And yet, those scientists who at once took to the canal interpretation, were by and large not astronomers. Until the 1880s Schiaparelli alone had seen them, and most astronomers did not believe that the fine lines existed, regardless of the interpretation. This is not as surprising as it might seem. Mars, even at a favourable opposition, is a small target, never more than about 25 arsec across (the Moon is about 1800 arcsec across). To see much at all on Mars's surface requires visual acuity, skill, experience, a good telescope, and good seeing (a clear steady atmosphere above the telescope). But astronomers had been alerted by Schiaparelli's observations, and by the next favourable opposition of Mars, in 1892, a few other astronomers had seen *canali*, and the more prominent *canali* had even been identified on maps earlier than 1877.

Nevertheless, the majority of astronomers could *not* see them, and did not believe in their existence, including the French astronomer E-M Antoniadi (1870-1944), a careful observer of Mars and an excellent draughtsman (Sagan & Pollack 1966). The minority that believed in *canali* divided into those that thought they were natural, and those that thought they were artificial. The natural interpretation was initially of water channels, joining one sea to another across the red continents. But within a few years of 1877 it was realised that the "seas" did not reflect light in the right sort of way. Then there were their changes in extent, shape, and contrast, for which no convincing explanations could be found in terms of seas. Moreover, the dark areas had structures within them, and some were crossed by *canali*. The interpretation of the dark areas by almost all astronomers therefore shifted away from oceans to either plant life, or minerals distinct from those in the bright areas (e.g. Lowell 1908, pp104-107). The *canali* were then thought to be some other natural feature such as strips of vegetation if the dark areas were plant life.

In the same way, those that thought the *canali* were artificial considered them at first to be water channels – canals – but as the interpretation of the dark areas shifted from seas to plant life then the *canali* were considered to be irrigated tracts of land, that what was seen was vegetation sustained by a thread of water too narrow to be seen. Indeed, the interpretation of the dark areas as plant life was preferred by this group, because the canal network could then be understood as an attempt by the martians to distribute their meagre water supply – hardly meagre if the dark areas were seas.

The idea that Mars was short of water stemmed from the interpretation of the bright, red tinted areas as deserts, and from the rapidity with which the polar caps advanced and retreated with the seasons, thus indicating that the caps were thin. Such a water shortage was seen as consistent with Mars's low surface gravity, due to its small size, which makes the escape of atmospheric constituents easier than on bigger planets like the Earth.

A dying world then, becoming desiccated. What more natural than for its inhabitants to build canals?

In the 1890s there emerged two powerful supporters of the view that the *canali* were canals. One was the US astronomer William Henry Pickering (1858-1938) who began to observe Mars in 1892. The other was another US astronomer, Percival Lowell (1855-1916), who in 1894 founded an observatory at Flagstaff in Arizona, mainly for him to study Mars. The site is appropriately named Mars Hill. Figure 7 shows a map of one hemisphere of Mars, drawn in 1905 (Lowell, 1908 facing p217).

> Figure 7    Mars in 1905, drawn by Percival Lowell. Note the canals. Note also the south polar hood of cloud (at the top).

Around this time the world of literature stirred in response to the possibility that there were intelligent martians. In 1898 appeared *The War of the Worlds* by the British writer Herbert George Wells (1866-1946, Wells 1898), a fine tale in which the martians look Earthwards and see
> "…… a morning star of hope, our own warmer planet, green with vegetation and grey with water, with a cloudy atmosphere eloquent of fertility, with glimpses through its drifting cloud wisps of populous country, and narrow, navy crowded seas" and we are destined to be invaded (Figure 8). This, and subsequent fiction involving Mars, kept alive well into the 20th century the belief among the general public that there was intelligent life on Mars.

> Figure 8    Martians, as depicted in an early edition of The War of the Worlds by H G Wells, which was first published in 1898. This drawing is by Warwick Goble.

By contrast, in the scientific community, such a belief, never widespread, was eroded by measurements that showed Mars is not as hospitable as had been thought at the end of the 19th century, but is generally more harsh than a dry Antarctic desert. Nevertheless, the belief in *canali* was sustained into the Space Age by a small minority of astronomers. Then, in 1965 the NASA spacecraft Mariner 4 flew by Mars – no *canali* were seen. Figure 9 shows three maps of the same area of Mars (Sagan and Fox 1975). Figure 9(a) shows a 19th century drawing by Schiaparelli, Figure 9(b) a drawing from 1929 by E M Antoniadi (1870-1944), and Figure 9(c) a map based on images from Mariner 9 and Earth based photography (Mariner 9 orbited Mars in 1971-2). Figure 9, and other image comparisons, make it clear that the *canali* do not exist. Comparison of Figure 9(a) and (b) is particularly telling. None of the *canali* in Figure 9(a) are present in Figure 9(b) – in superior seeing conditions Antoniadi found the *canali* to be loosely aligned spots and streaks.

> Figure 9    Three maps of the same area of Mars, about 4400 km across (a) A drawing by Schiaparelli in the 19th century (b) A drawing by Antoniadi from 1929 (c) A map of Mars based on images from the 1971-2 Mars orbiter Mariner 9 and Earth based photography.

What then are we to make of Lowell (1908, p215)
> "…… not only do the observations…(on the canals)…we have scanned lead us to the conclusion that Mars at this moment is inhabited, but they land us at the further one that these denizens are of an order whose acquaintance was worth the making."

Or even of Schiaparelli (1894)
> "It is not necessary to suppose them to be the work of intelligent beings, and notwithstanding the almost geometrical appearance of all of their system, we are now inclined to believe them to be produced by the evolution of the planet, just as on the Earth we have the English Channel and the Channel of Mozambique."

The explanation seems to be that the *canali* were indeed evidence of intelligent life, but, as the US astronomer Carl Sagan (1934-1996) put it, the intelligence was at the eyepiece end of the telescope. The human mind, straining to interpret elusive detail at the limit of perception, invented narrow

IJA Mars pre space age + Figs  BWJ   Version 05Nov2008      10 of 28                    13/11/2008  19:30 UT

linear features that are simply not there. At best there are roughly aligned spots and streaks. The canals stand not as a chronicle of Mars but as a monument to the subtleties of human visual perception.

**The green and red planet**

Aside from the *canali*, what was our picture of Mars in the years before Mariner 4 flew by in July 1965? Accounts are to be found, for example, in Sheehan (1996), Jackson & Moore (1965), Slipher (1962), Strughold (1954), and Vaucouleurs (1950). Here is a summary of the main points.

*The surface*

One of the best pre space age maps of Mars is shown in Figure 10. Such maps are based on photographs through Earth based telescopes and also on visual observations. Photography had the advantage that it gave accurate shapes and that very low contrast features were revealed. Visual observations had the advantage that glimpses of Mars during moments of very good seeing yielded detail that was beyond the reach of photography, because during typical exposure times turbulence in the Earth's atmosphere blurred fine detail.

> Figure 10    The International Astronomical Union albedo map of Mars, current in the early 1960s. South is at the top. Presumably the features shown are seasonally averaged. Note that the polar regions are not included.

Maps of Mars at this time showed albedo features, not topography. The only reliable topographic data were from radar. These data were of low spatial resolution, though they did indicate altitude differences up to 16 km between different areas of Mars. In fact, the altitude range is greater than this, though only because of local features.

What of the seasonal and non-seasonal changes in the albedo features? There was evidence that in the spring hemisphere, in which the polar cap was consequently retreating, the dark areas became even darker. There was even some evidence of a "wave of darkening" spreading from the waning polar cap towards the equator. This supported the long held belief that the dark areas were plant life, being revived in the spring by the rising temperatures and by moisture released from the polar cap. Non-seasonal changes were then due to changes in the weather from year to year.

What of the astronomers who believed that the dark areas were distinct from the bright areas in being of a different mineralogical composition? One of these was the US astronomer Dean Benjamin McLaughlin (1901-1965) who, in the 1950s, proposed that the dark areas were ash from still-active volcanoes, placed in semi-permanent patterns by the prevailing winds (Veverka & Sagan 1974). It had also been suggested that the dark areas contained substances that darkened as they absorbed water, thus accounting for the increased darkness of the dark areas in spring (Vaucouleurs 1950, pp70-72). Another suggestion was that the dark areas darkened in spring as the known light patches within the dark areas were filled with new dark spots.

The bright areas were regarded as dusty deserts by almost all astronomers. Some of the variation in the size and shape of the dark areas could be due to a battle between the growth of plants and the encroaching desert.

There was a comparable degree of agreement that the polar caps consisted of condensed water in the form of snow or frost. Their rapid seasonal advances and retreats indicated a thickness in these seasonal caps of no more than 100-200 mm (the residual caps at each pole could be much thicker).

Temperatures at the martian surface had been measured from Earth by radiometry and spectrometry. Near midday in the martian tropics the surface could reach a high of about 280 K (7°C), but near sunrise the surface temperatures were typically 228 K.

*The atmosphere*

These large diurnal swings in temperature showed that the martian atmosphere was a good deal less effective than that of the Earth in blocking planetary radiation to space, and thus allowed prodigious cooling of the martian surface. It followed that the column mass of the atmosphere, and the surface pressure, were a good deal less than on Earth. Detailed analysis of the solar radiation scattered by Mars, plus some rather wobbly assumptions about the interaction of solar radiation with the surface and atmosphere, led to estimates of the atmospheric pressure at the martian surface in the range 80-120 millibars with some preference for the lower values. This is considerably less than the 1000 millibars or so at the Earth's surface The column masses are more similar because the surface gravity on Mars is only 38% of that on the Earth: Mars, $0.21\text{-}0.32 \times 10^4$ kg m$^{-2}$; Earth, $1.03 \times 10^4$ kg m$^{-2}$.

The composition of the atmosphere had been investigated by means of spectrometry. Solar radiation scattered by Mars and reaching the Earth will have both atmospheric and surface features impressed on it. The atmospheric signatures can be distinguished (e.g. they are narrower than the surface signatures), and thus atmospheric gases can be identified. By 1965 it had been known for some years that $CO_2$ was present in the martian atmosphere, and that it accounted for only a few millibars of the 100 millibar or so total pressure. Water vapour had been detected at the detection limit, a few hundredths of a millibar, making it clear that the martian atmosphere is far drier than the atmosphere of the Earth.

The bulk of the 100 or so millibars thus remained unaccounted for. A widely held view was that, as in the Earth's atmosphere, $N_2$ was the predominant component. There was little hope of detecting $N_2$ from the Earth. First, over the wavelength ranges that dominate solar radiation (near UV, visible, near IR) $N_2$ has a weaker spectral signature than $CO_2$ and water vapour. Second, the copious amount in the Earth's atmosphere would mask any martian signal. This is also the case for water vapour. However, by making observations at high altitude desert sites the amount of terrestrial water vapour above the telescope was greatly reduced. Also the Doppler shifts induced by the motion of Mars with respect to the Earth slightly separated the martian and terrestrial spectral lines. This separation had aided the identification of martian $CO_2$. However, $CO_2$ is only a trace in our atmosphere, contributing only about 0.37 millibars, so the martian signal was not difficult to discern.

$O_2$ was as difficult to detect as $N_2$, and for the same reasons. However, it seemed unlikely that any biosphere had been extensive enough to generate much $O_2$ (by photosynthesis), and therefore most astronomers thought that $O_2$ accounted for much less than a millibar of the martian atmosphere.

Three types of cloud had been recorded. The blue clouds are high altitude thin hazes, called blue because they are best seen in blue light. They were widely regarded as composed of tiny crystals of water ice, though tiny $CO_2$ crystals were an alternative possibility.

The white clouds (usually sparse) were also widely regarded as composed of tiny crystals of water ice, rather like cirrus in our atmosphere. Nearly all white clouds are small, and some sites on the surface have such clouds relatively often. On the Earth mountain peaks and basins are favoured cloud sites, and the same was thought to be the case for Mars. The largest white cloud by far is the sinister sounding polar hood (Figure 7), a cloud that spreads across the polar regions in each hemisphere during the autumn, thus hiding from view the major phase of seasonal growth of the polar cap from the cloud from which the winter snows fell. At their maximum extent the polar hoods spread about half way to the equator. Only in spring does the hood disappear, to reveal a greatly enlarged polar cap.

The yellow clouds were widely regarded as clouds of desert dust raised by strong winds. These cloud are often extensive, and sometimes obscure the whole martian globe for several weeks.

Such, in broad outline, was the Earth-based view of Mars in 1965, a view based on observations from a distance of never less than about 55 million km, obtained through the Earth's turbid turbulent atmosphere. Then, Mariner 4 flew by Mars.

**Mariners 4, 6, and 7**

On 15 July 1965 the NASA spacecraft Mariner 4 flew past Mars at a minimum distance of only 9800 km. There were two major surprises. First, all 22 of the images sent to Earth showed impact craters (Figure 11), the result of small bodies colliding with the martian surface at very high speeds. The heavily cratered (and hence ancient) surfaces indicate lack of geological activity and lack of extensive weathering by the atmosphere and by condensates of water, which would have erased these craters in a fraction of the age of Mars. Second, as Mariner 4 passed beyond Mars it moved behind the planet as viewed from the Earth. As it did so the radio transmissions from the spacecraft passed through the martian atmosphere, and the changes induced by the atmosphere enabled the surface pressure of the atmosphere to be determined. It was not the 100 millibars or so that had been confidently expected, but a mere 6 millibars! It followed that nearly all of the atmosphere could be accounted for by the $CO_2$ that had previously been detected from the Earth, and we now know that 95% of the molecules in the atmosphere are $CO_2$, and that $N_2$ and Ar accounts for almost all the rest (the fraction by number of molecules is the same as the fractional contribution to the total surface pressure).

> Figure 11    One of the images sent to Earth by Mariner 4 during its flyby of Mars in July 1965. This image is centred 26.8° south of the equator and is 267 km across. (NASA, Mariner 4 frame 10D)

A pressure of 6 millibars is close to the triple pressure of water (6.10 millibars), below which water cannot exist in a stable liquid phase at any temperature. All life on Earth requires liquid water, so its likely scarcity at the martian surface dealt a heavy blow to the prospect of finding life on Mars. The blow was reinforced by the, at best, low $O_2$ content of the atmosphere, indicating the absence of copious oxygenic photosynthesis. It was also reinforced by the craters, making Mars seem more like the sterile Moon than the fertile Earth.

The dominance of $CO_2$ in the atmosphere led to the revival of an old idea that the polar caps were not made of water but of $CO_2$, which also forms a white snow. This idea gained support from the next two successful missions to Mars, the flyby of Mariner 6 in July 1969 and the flyby of Mariner 7 a few days later in August 1969. The temperature of the south polar cap was measured and found to correspond to the solid-gas phase boundary of $CO_2$ at a pressure of a few millibars. This provided strong evidence that a polar cap of $CO_2$ was roughly in equilibrium with the $CO_2$ atmosphere. Subsequent studies have confirmed that the seasonal cap at both poles is indeed predominantly $CO_2$ snow and frost, but that this overlies a permanent cap mainly composed of dusty water ice at the North Pole, and dusty $CO_2$ ice at the colder South Pole, perhaps underlain by dusty water ice.

**Mariner 9**

Mariners 4, 6, and 7, had imaged only the southern hemisphere of Mars. On 14 November 1971 the spacecraft Mariner 9 was placed in orbit around Mars. It mapped the *whole* planet and marks the start of the contemporary era of martian exploration. Figure 12 shows the albedo-topographic map of Mars established by Mariner 9. You can that the heavily impact cratered ancient terrain is confined to the southerly hemisphere of Mars – south of a line tilted at nearly 40° with respect to the equator. North of this line there are vast plains, volcanic domes, shield volcanoes, rift valleys, and many other features, all indicating geological activity extending to within the past few million years in some places.

      Figure 12      The albedo-topographic map of Mars established by Mariner 9 (the polar regions are not shown). North is at the top.

Mariner 9 confirmed that the bright areas are arid deserts of sand and dust composed of basaltic minerals rich in iron (hence the red tint) and magnesium. Clays are also present resulting from the aqueous alteration of silicates. Mariner 9 found that the dark areas are also sand and dust composed of basaltic minerals rich in iron and magnesium, plus underlying basaltic rock exposed here and there. The dark material is thought to give rise to the bright material through various physical and chemical processes.

Mariner 9 solved the mystery of the changes in the dark areas. It was not long before it was shown that the sand and dust in the bright and dark areas is mobilized by martian winds, through the interaction of winds with small scale topography, such as impact craters. Seasonal and non-seasonal changes in wind speed and direction cause the seasonal and non-seasonal changes respectively. The streakiness of the surface, particularly in the dark areas is also due to the winds. See Veverka and Sagan (1974).

The most exciting discovery made by Mariner 9, certainly as far as life on Mars is concerned, is many features that seem to have been carved by the flow of liquid water. Figure 13 shows three types of channel. These are considerably more common in the ancient southerly hemisphere and indicate that Mars was much warmer and wetter in the first 1000 million years or so of the 4600 million years of martian history. Several other types of feature indicating the presence of liquid were discovered. More recently, gulleys have been discovered that indicate much more recent flows, perhaps even occurring today (Figure 2). It is now widely believed that liquid water lies at depths of no more than a few km, and at more shallow depths in some locations.

      Figure 13      Channels on Mars that seem to have been carved by the flow of liquid water, early in martian history. (a) The outflow channel at the head of Simud Vallis. Frame width about 300 km. (b) The fretted channel Nirgal Vallis. Frame width about 160 km. (c) A valley

network, common in the southerly hemisphere of Mars. Frame width about 130 km. (NASA/JPL)

Since Mariner 9 there have been many spacecraft missions to Mars, including several landers. We now know a lot more about our planetary neighbour, but the Mariner 9 discoveries outlined here still stand. See Jones (2007) for an account of what we know about Mars today.

**A martian biosphere?**

You have seen that the belief that Mars was inhabited, if only by vegetation, persisted up to the flyby of Mariner 4 in 1965. Then Mariner 4 revealed an ancient cratered landscape more akin to the Moon, and a thin atmosphere dominated by $CO_2$. The belief in martian lifeforms waned rapidly.

You then saw that the arrival of Mariner 9 in 1971 restored the picture of Mars as an interesting, distinct place, and thet there is ample evidence that Mars was warmer and wetter early in its history. Could there be fossils left from that time? We don't know. A sample return mission to the ancient cratered areas is needed. A joint ESA-NASA sample return mission is being developed, which, if approved, would see a launch in the next decade (ESA 2007).

Could there be a deep subsurface biosphere? Again, we don't know. None of the landers have found any evidence of life. We await a Mars lander devoted to astrobiology, such as the astrobiology field laboratory under consideration by NASA for the next decade (NASA 2007). However, it is certainly the case that the question of life on Mars is still open.

**Summary**

Our picture of Mars today is not only much more complete than that before Mariner 4 in 1965, in several ways it is also quite different. This summary focuses on aspects particularly relevant to life on Mars. The *italicized text* is the picture in the decade before the flyby of Mariner 4 in 1965, based on difficult-to-make observations from the Earth. The non-italicized text is the current picture, based on data from several orbiters and five landers. These data place the current view on a firm foundation. The current view, as summarised here, is not very different from that prevalent in the early 1970s.

Bright areas

*Arid deserts of sand and dust, perhaps tinted by iron-rich minerals.* Arid deserts of sand and dust composed of basaltic minerals rich in iron (hence the red tint) and magnesium, mixed with clays resulting from the aqueous alteration of silicates.

Dark areas

*Majority view – some kind of primitive living material, "vegetation". Minority view – some kind of mineral, distinct from the bright areas.* Sand and dust composed of basaltic minerals rich in iron and magnesium, plus underlying basaltic rock exposed here and there. Though darker than the bright areas, there is a red tint. The dark material is thought to give rise to the bright material through various physical and chemical processes (including aqueous alteration).

Seasonal changes in the dark areas

*The seasonal darkening in the spring hemisphere is caused by water released by the receding polar cap in that hemisphere. This rejuvenates the "vegetation". If the dark areas are non-organic, the darkening might result from the dampening of hygroscopic salts, or the spring infill of light patches with dark spots.* Seasonal changes in the winds cause the changes in the dark areas through the interaction of wind speeds and directions with small scale topography.

Non-seasonal changes in the dark areas

*These are due to the spread of "vegetation" in favourable times, lasting longer than a martian year, and its retreat in unfavourable times.* The cause is changes in the winds on a timescale greater than a martian year.

Atmospheric composition

| mean surface values | pre space age/ mbar | current/ mbar |
|---|---|---|
| total pressure | 80-100 | 5.60 |
| $CO_2$ | a few | 5.32 |
| $N_2$ | the rest[a] | 0.15 |
| Ar | ——— | 0.10 |
| $O_2$ | $\ll 1$ | 0.007 |
| water vapour | 0.01-0.1 | ~ 0.0006[b] |

a  assumed value
b  close to the saturation value

Clouds

| type of cloud | pre space age | current |
|---|---|---|
| "blue" | high altitude thin hazes made of tiny crystals of water ice, or perhaps $CO_2$ ice | high altitude thin hazes made of tiny crystals of $CO_2$ ice |
| white | particles of water ice | tiny crystals of water ice; tiny crystals of $CO_2$ ice over the winter pole |
| yellow | local to global, desert dust raised by strong winds | local to global, predominantly wind raised dust from the bright areas |

Surface temperatures (sample)

| pre space age | current |
|---|---|
| in the tropics: –45 °C just before dawn, up to about 7°C at noon | – 130 °C over the winter pole, up to around 20 °C at noon in the tropics |

Polar caps

*Condensed water in the form of snow or frost. Their rapid seasonal advances and retreats indicates a thickness in these seasonal caps of no more than 100-200 mm (the residual caps at each pole could be much thicker).* The seasonal caps consist of $CO_2$ ice. The residual southern cap also consists of $CO_2$ ice, presumed to be underlain by dusty water ice. By contrast, the residual northern cap is water ice (plus some dust).

Topography

*Altitude differences up to 16 km between different areas of Mars seen by radar. The frequent occurrence of small white clouds at certain locations indicate the presence of mountain peaks.* Nearly all of the surface lies within an altitude range of 12 km, though a few volcanic peaks reach to over 20km above the mean surface level. Heavily impact cratered terrain is confined to the southerly hemisphere of Mars. North of this line there are vast plains, volcanic domes, shield volcanoes, rift valleys, and many other features indicating geological activity extending to within the past few million years in some cases.

Water and life (including the canals)

*Though atmospheric water vapour accounts for only a few hundredths of a millibar, if the dark areas are plant life the release of water from the receding polar cap is sufficient to reactivate the "vegetation", causing a spring darkening. Higher forms of life, even intelligent life once held to have constructed canals to distribute the meagre water supply from the polar caps, are ruled out – the canals do not exist and the thin atmosphere, almost devoid of oxygen is unsuitable for higher*

*forms of life as we know it*. There is ample evidence that water flowed on Mars in its distant past, and it is possible, even likely, that fossils of lifeforms survive from that time. Liquid water seems able to make brief appearances on Mars today, so a subsurface biosphere, presumably microbial, cannot be ruled out.

Our view of Mars has changed considerably since the dawn of the Space Age, but it is just as fascinating a planet as before. Fossils are a possibility, and even extant microbial life might be present. We look to future space missions resolve these issues.

Wells, H.G. (1898). *The war of the worlds*. Many editions, Chapter 1.

**FIGURES – for higher definition see Int. J. Astrobiol., 7, 143-155 (2008)**

Figure 1

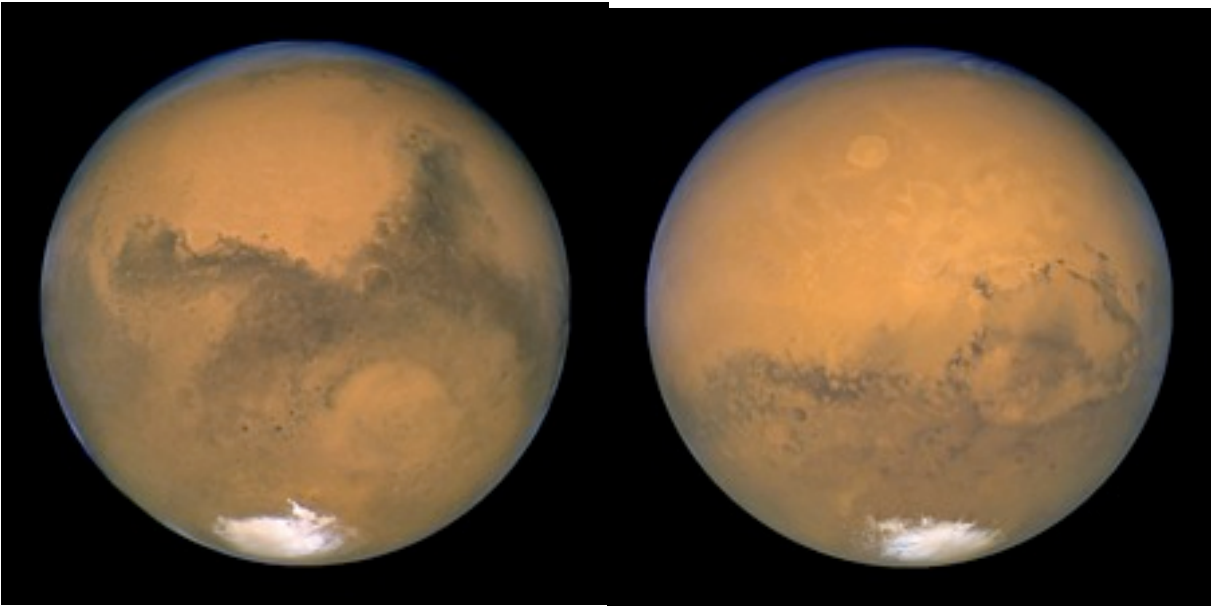

Figure 2

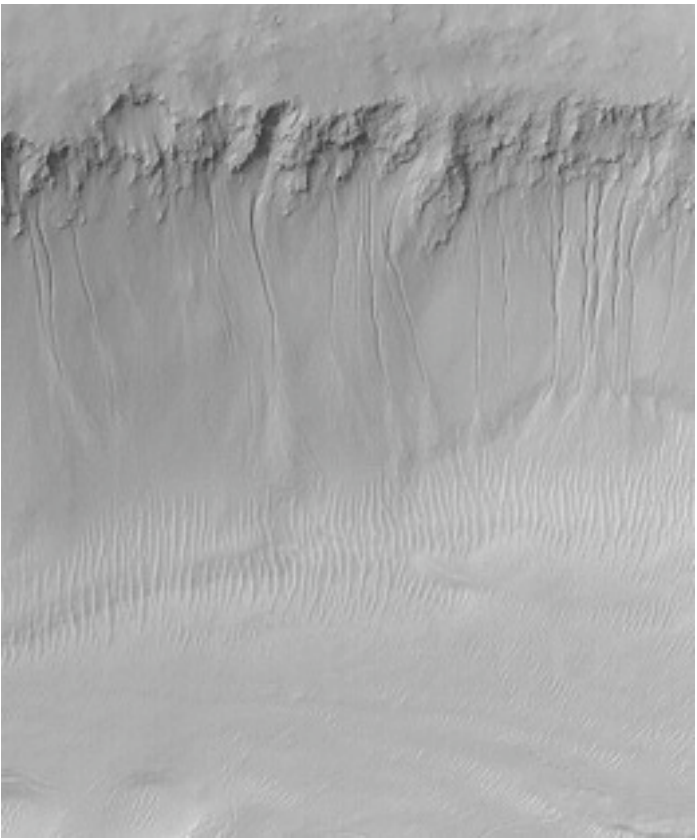

Figure 3

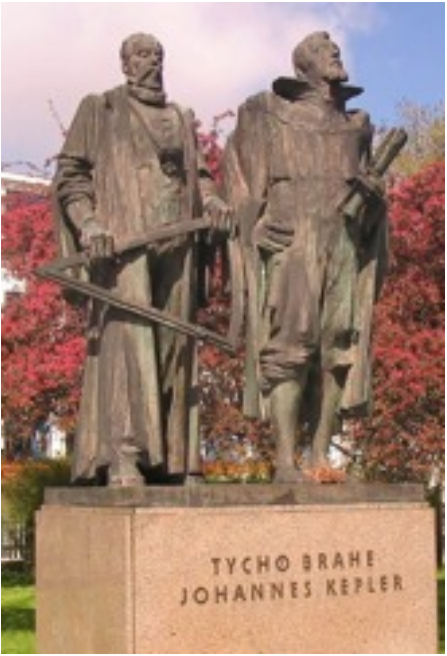

Figure 4

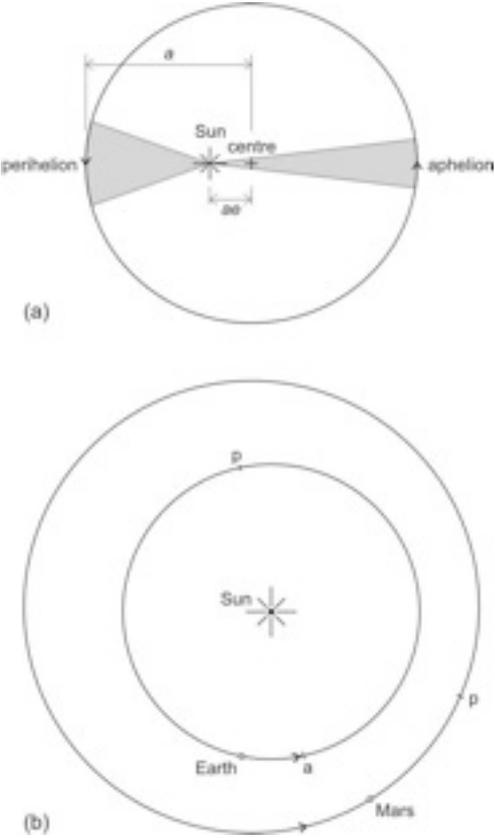

Figure 5

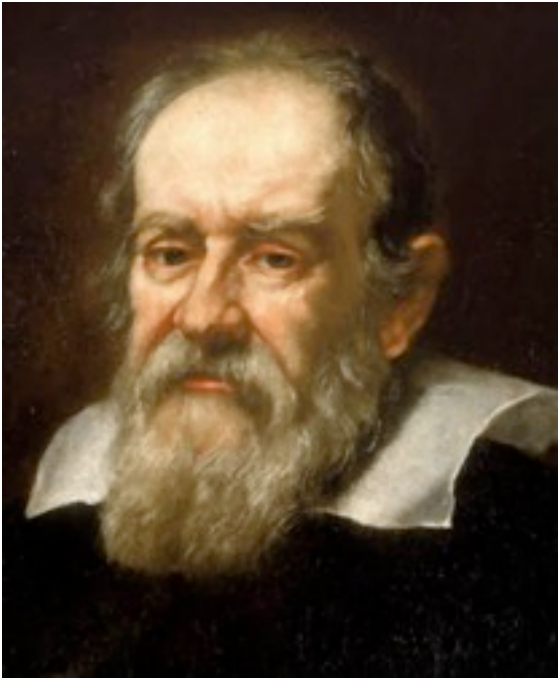

(a)

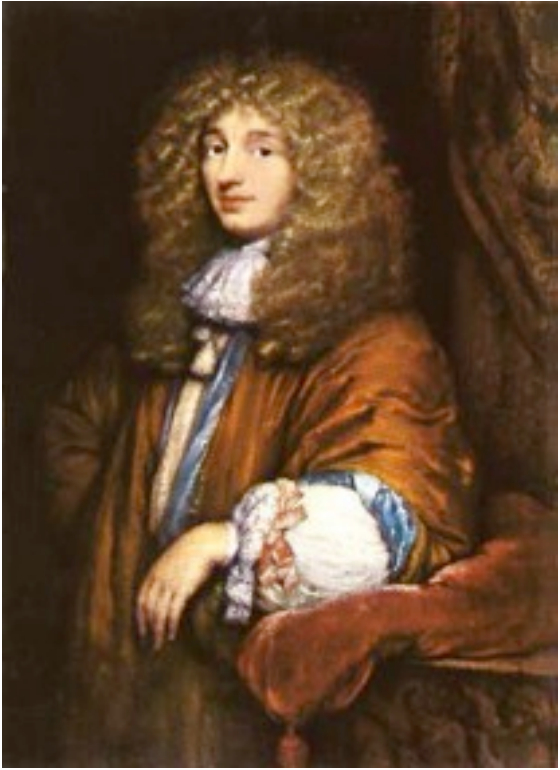

(b)

Figure 6

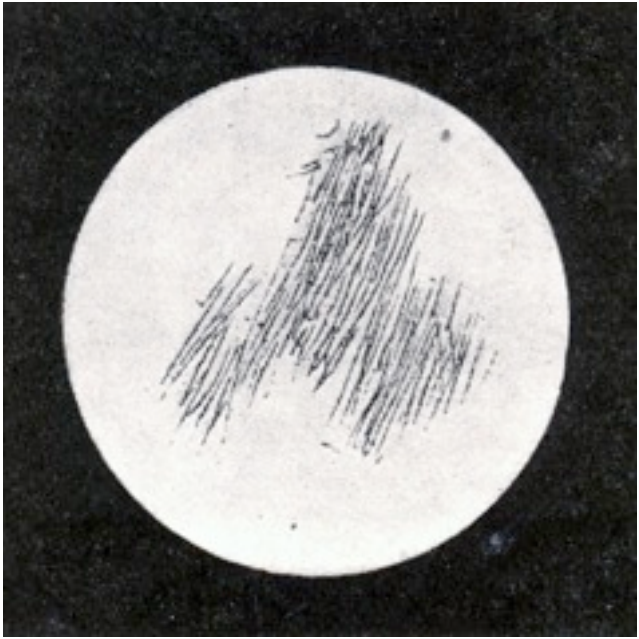

Figure 7

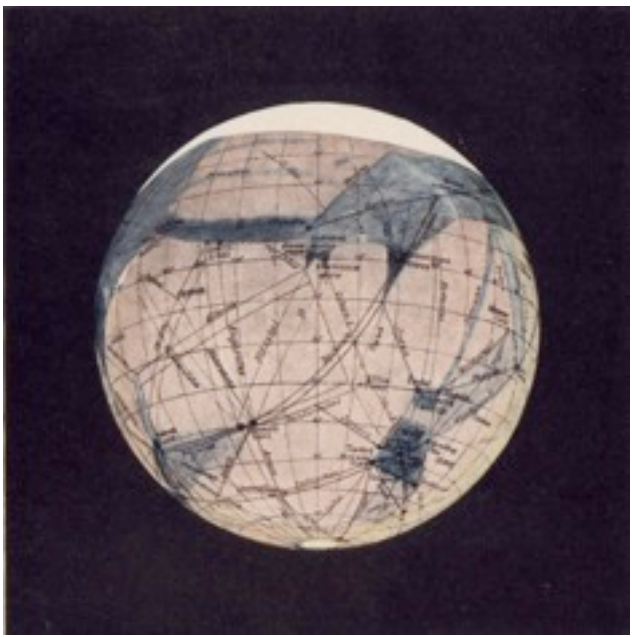

Figure 8

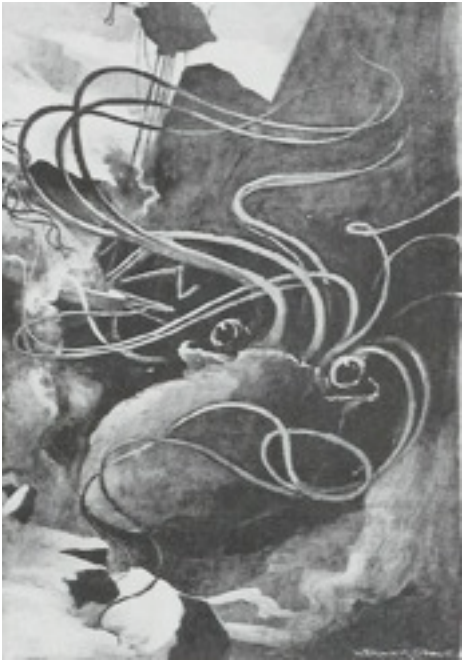

Figure 9

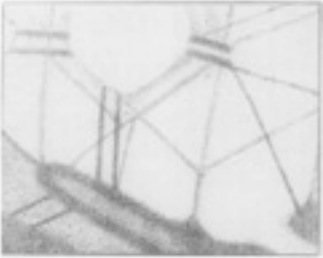
(a)

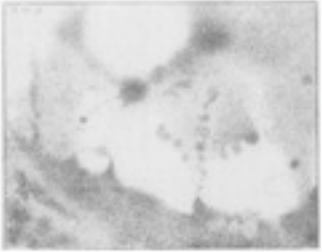
(b)

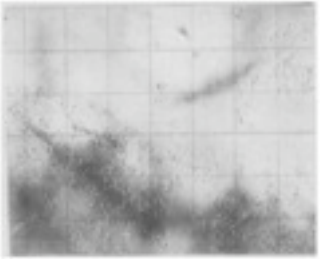
(c)

Figure 10

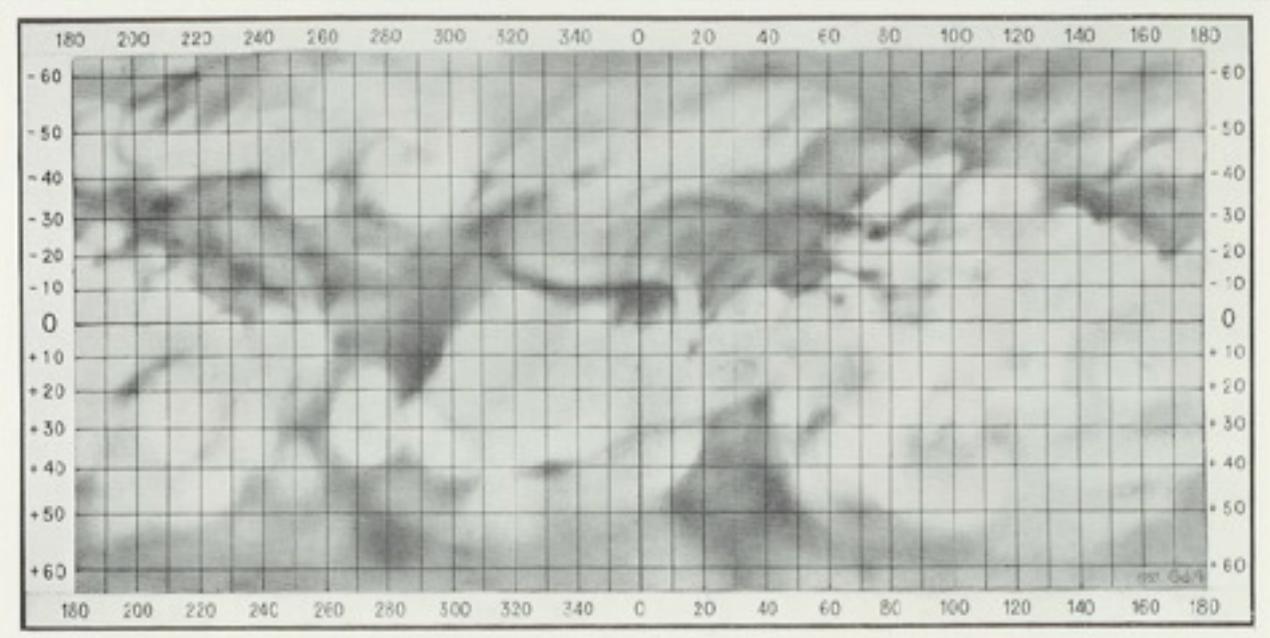

Figure 11

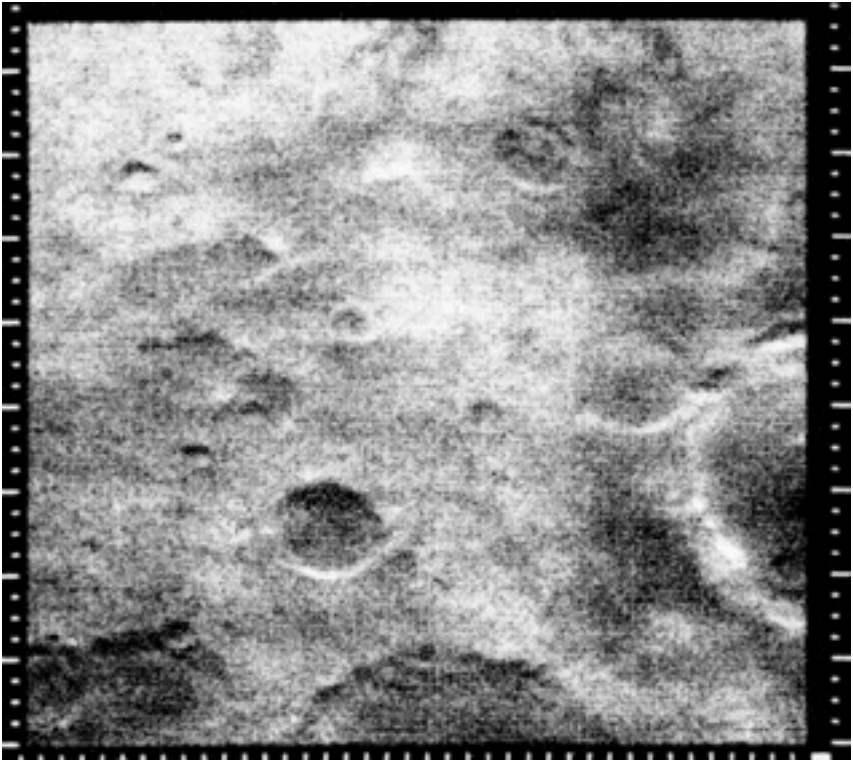

Figure 12

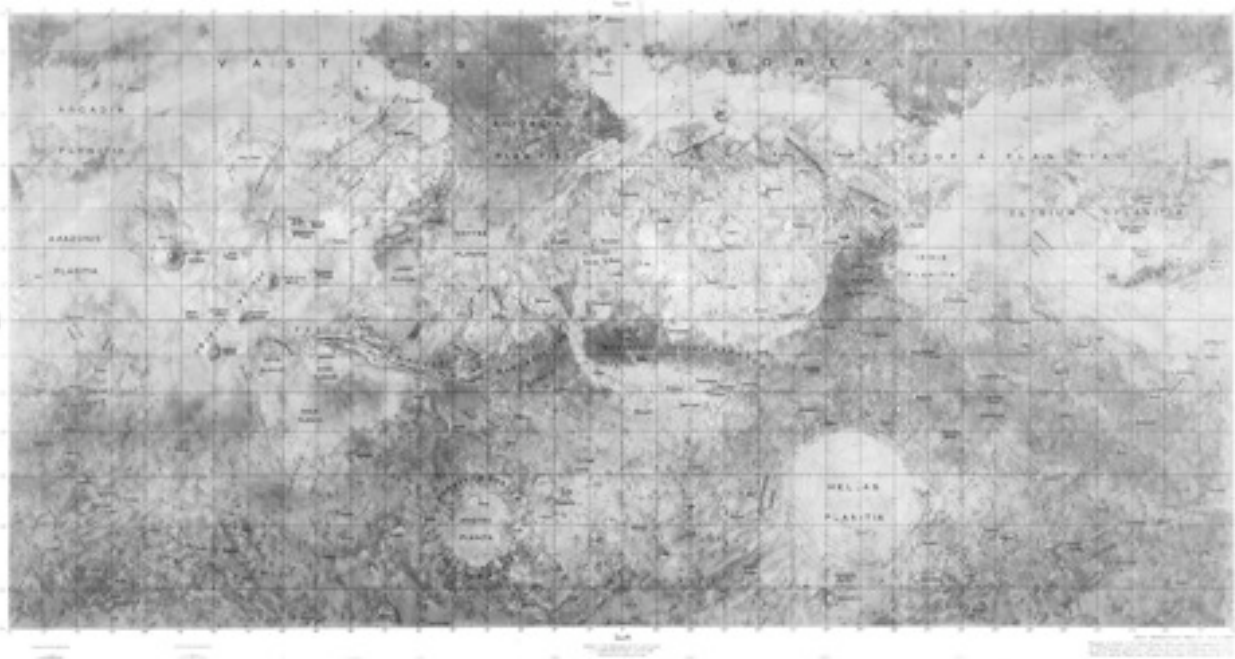

Figure 13

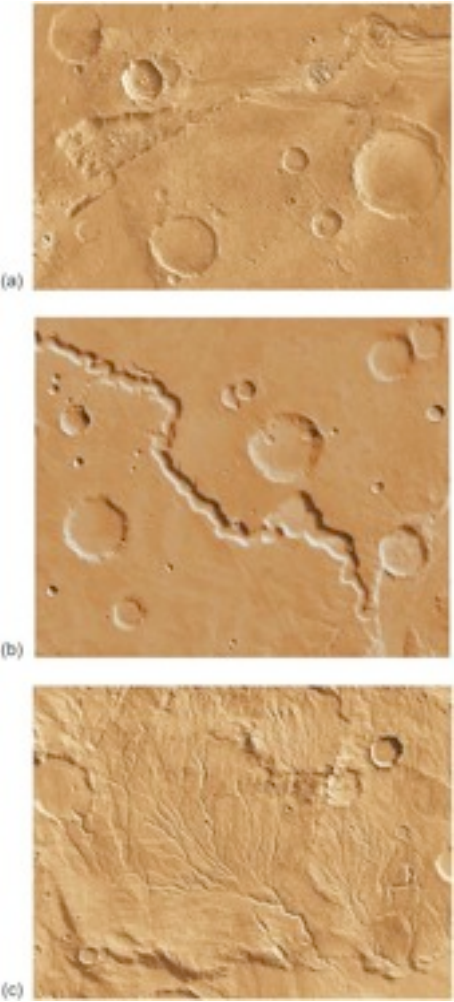